\documentstyle[twoside,fleqn,espcrc2,epsf,rotate]{article}

\newcommand{\AmS}{{\protect\the\textfont2
  A\kern-.1667em\lower.5ex\hbox{M}\kern-.125emS}}

\title{On the $I=2$ channel $\pi$-$\pi$ interaction in the chiral limit
       \thanks{Supported in part by NSF PHY-9700502, 
                                    and by FWF P10468-PHY.}}

\author{H. R. Fiebig\address{Physics Department, 
                            FIU-University Park, 
                            Miami, Florida 33199, USA},
        H. Markum\address{Institut f\"{u}r Kernphysik, 
                          Technische Universit\"{a}t Wien, 
                          A-1040 Vienna, Austria},
        A. Mih\'{a}ly\address{Department of Theoretical Physics, 
                              Lajos Kossuth University,
                              H-4010 Debrecen, Hungary},
        and
        K. Rabitsch$^{\mbox{\scriptsize b}}$}

\begin{document}

\begin{abstract}
An approximate local potential for the residual $\pi^+$--$\,\pi^+$ interaction
is computed. We use an $O(a^2)$ improved action on a coarse $9^3\times 13$
lattice with $a\approx 0.4$fm. The results present a continuation
of previous work: Increasing the number of gauge configurations and 
quark propagators we attempt extrapolation of the $\pi^+$--$\,\pi^+$ potential
to the chiral limit.
\end{abstract}

\maketitle

\section{INTRODUCTION}

The use of improved lattice actions allows to work with lattice volumes
large enough to accommodate systems of two hadrons, with manageable 
computational effort. We take advantage of this opportunity to study
the residual effective interaction of two pseudo-scalar mesons on the
lattice. 

This paper is a report on the current status of this project. An earlier
exploratory study \cite{Fie98b} was made with a smaller number
of gauge configurations and two, somewhat large, quark masses.
The new results presented here are based on 208 configurations and 6
values of quark masses, with otherwise unchanged lattice parameters. 
The current simulation allows extrapolation of the extracted
$\pi$-$\pi$ potential to the chiral limit
and a comparison of lattice-based scattering phase
shifts, computed with the potential, to experimental results
\cite{Fro77,Mae74} in the isospin $I=2$ channel.

\section{LATTICE PARAMETERS}

An $L^3\times T = 9^3\times 13$ lattice was used with an $O(a^2)$ tree-level
and tadpole improved action based on next-nearest neighbor couplings
\cite{Ham83,Egu84}. At $\beta=6.2$, in the conventions of \cite{Fie96d},
the corresponding lattice constant is $a\approx 0.4$fm, or 
$a^{-1}\approx 500$MeV. We have used $N_U=208$ quenched gauge configurations.
The hopping parameters for Wilson fermions were set to 
$\kappa^{-1}=5.720, 5.804, 5.888, 5.972, 6.056, 6.140$.
The critical value for $\kappa^{-1}$ is $\approx 5.5$.
Quark propagator matrix elements were computed using a random-source 
technique \cite{Can97} with $N_R=8$ Gaussian sources.

\section{METHOD}

An outline of the theoretical framework may be found in \cite{Fie98b,Can97},
we here mention only the essential points. Suitable operators for the 
$\pi^{+}$--$\pi^{+}$ system, having isospin $I=2$, are
\begin{equation}
\Phi_{\vec{p}}(t)=\phi_{-\vec{p}}(t)\,\phi_{+\vec{p}}(t) \,,
\end{equation}
where
\begin{equation}
\phi_{\vec{p}}(t)=L^{-3}\sum_{\vec{x}}\,e^{i\vec{p}\cdot\vec{x}}
\bar{\psi}^{\mbox{\scriptsize d}}(\vec{x},t)\,\gamma_5
\psi^{\mbox{\scriptsize u}}(\vec{x},t)
\end{equation}
describes single mesons with lattice momenta $\vec{p}$.
The correlation matrix for the $\pi^{+}$--$\pi^{+}$ system
\begin{equation}
C_{\vec{p}\,\vec{q}} =
\langle\Phi^{\dagger}_{\vec{p}}(t)\,\Phi_{\vec{q}}(t_0)\rangle =
\bar{C}_{\vec{p}\,\vec{q}}+C_{I,\vec{p}\,\vec{q}}
\end{equation}
is a sum of a free, $\bar{C}$, and a residual-interaction contribution, 
$C_{\mbox{\scriptsize I}}$.
The free $\pi^{+}$--$\pi^{+}$ correlator $\bar{C}$ is diagonal in $p,q$.
We also implement link variable fuzzing and operator 
smearing \cite{Alb87a,Ale94} at the sink.

We define an effective interaction through
\begin{equation}
H_{\mbox{\scriptsize I}} = -\frac{\partial}{\partial t} \ln 
(\bar{C}^{\,-1/2}\,C\phantom{x}\bar{C}^{\,-1/2}) \ .
\end{equation}
Matrix elements of $H_{\mbox{\scriptsize I}}$ are obtained from linear
fits to the logarithm of the eigenvalues of the correlators $\bar{C}$
and $C$.
At this point, only the diagonal elements of $C$ were utilized.
Momentum-space matrix elements 
$(\vec{p}\,|H_{\mbox{\scriptsize I}}|\vec{q}\,)$ 
are computed in a truncated basis of small lattice momenta.
The Fourier transform to coordinate space 
contains a local potential
\begin{equation}
V(\vec{r}\,) =
\sum_{\vec{q}} e^{-2i\vec{q}\cdot\vec{r}}
(-\vec{q}\,|H_{\mbox{\scriptsize I}}|+\vec{q}\,) \,.
\end{equation}
Its $s$-wave projection ($\ell = 0$) makes use of only the irreducible
representation $A_1$ of the lattice symmetry group $O(3,{\cal Z})$.
In terms of the corresponding reduced matrix elements we have
\begin{equation}
V(r) =
\sum_{q} j_0(2qr) (q|H_{\mbox{\scriptsize I}}^{(A_1)}|q) \ ,
\label{sVloc}\end{equation}
where $j_0$ is a spherical Bessel function.

\section{RESULTS}

\begin{figure} \centering
\framebox{\epsfxsize=55mm\rotate[l]{\epsfbox{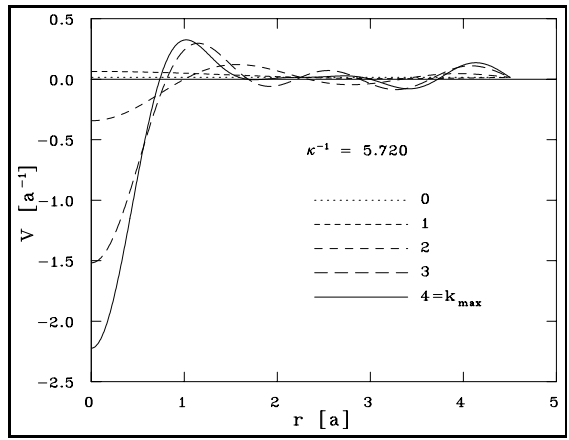}}}
\caption{Local potentials (\protect\ref{sVloc}) for the residual $\pi$-$\pi$ 
interaction at $\kappa^{-1}=5.720$. Results using truncated
correlation matrices of sizes one-through-five
($k_{\mbox{\scriptsize max}}=0,1,2,3,4$) give a hint at the convergence
behavior of the Fourier series.}
\label{fig1} \end{figure}

Potentials according to (\ref{sVloc}) for one selected value of
$\kappa$ are shown in Fig.~\ref{fig1}. The sum over
(on-axis) momenta $q=\frac{2\pi}{L} k$ was truncated at increasing
$k_{\mbox{\scriptsize max}}=0,1,2,3,4$, respectively.
A detailed error analysis is in progress, however, errors appear to
increase with $k$. The results presented below are for the truncation at
$k_{\mbox{\scriptsize max}}=3$.

Chiral extrapolation of the potential was done by linear fits of
$V(r)$ versus $m^2$ for a fine (plot-grade)
mesh of values $r$, fixed one at a time.
Using sets of 3 through 6 data points, corresponding to the smallest
available values of $m^2$, gives very similar results.
The subsequent analysis was performed with 3 data points.

\begin{figure}[t] \centering
\framebox{\epsfysize=79mm\epsfbox{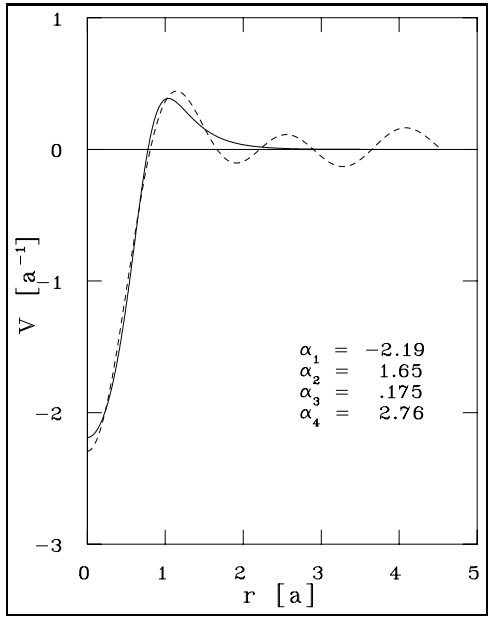}}
\caption{Result $V(r)$ of the chiral extrapolation shown by the dashed line.
A parametric fit with $V^{(\alpha)}(r)$, see (\protect\ref{Valpha}),
gives the solid line.}
\label{fig2} \end{figure}

The extrapolated potential $V(r)$ is shown in Fig.~\ref{fig2} as a dashed
line. The oscillations of $V(r)$
in the region $r>2$ are due to the Fourier transform of the truncated
momentum sums. The wave length is indicative of the lattice resolution
at the current truncation.

A parametric fit to $V(r)$ with 
\begin{equation}
V^{(\alpha)}(r) = \alpha_1\frac{1-\alpha_2 r^{\alpha_5}}
{1+\alpha_3 r^{\alpha_5+1}e^{\alpha_4 r}} + \alpha_0 \,,
\label{Valpha}\end{equation}
at $\alpha_5=2$ fixed, was applied to the extrapolated potential.
The result is shown in Fig.~\ref{fig2} as a solid line.
It suggests attraction at short distances followed by a repulsive
barrier.

\begin{figure}[t] \centering
\framebox{\epsfysize=79mm{\epsfbox{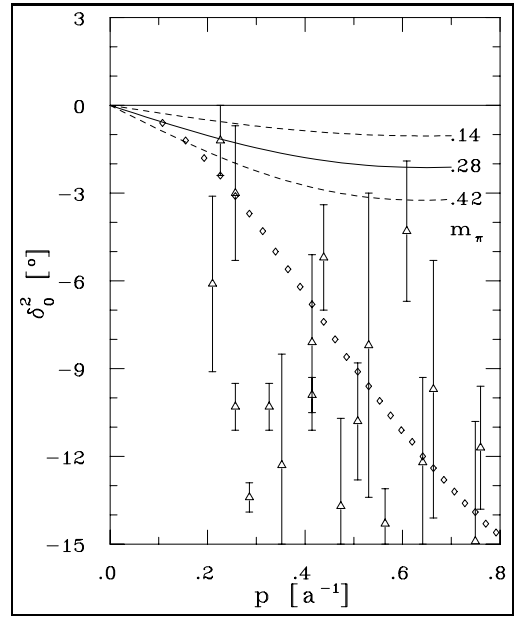}}}
\caption{Scattering phase shifts $\delta_{0}^{2}(p)$ 
computed with the
parametrization $V^{(\alpha)}$ of the chirally extrapolated potential.
Results for three (ad hoc) values of $m_\pi$ are shown.
(Convert units with $a=0.4$fm and $a^{-1}=500$MeV.) The experimental
pion mass corresponds to $m_\pi=0.28$. Experimental data are from
\protect\cite{Fro77} ($\Diamond$) and \protect\cite{Mae74} ($\triangle$).}
\label{fig3} \end{figure}

We have used $V^{(\alpha)}(r)$ in a Schr{\"o}dinger
equation to calculate s-wave scattering phase shifts
$\delta_{\ell=0}^{I=2}(p)$,
see Fig.~\ref{fig3}. The pion mass was set to multiples of the
experimental value, corresponding to $m_\pi=0.28$ in units of $a^{-1}$.
The repulsive nature of the phase shifts is due to the hump of
$V^{(\alpha)}(r)$ around, and extending beyond, $r\approx 1$,
see Fig.~\ref{fig2}.
The data points in Fig.~\ref{fig3} are experimental results compiled
from \cite{Fro77,Mae74}.

\section{CONCLUSION}

Scattering phase shifts for the $I=2$ channel $\pi$--$\pi$ system
were computed from lattice QCD by way of extracting a non-relativistic
potential. 

Since at this point an error analysis is pending,
particularly systematic, errors are unknown.
The range of the extracted potential is short compared to
the current spatial resolution. The latter is determined by 
the somewhat large value of the lattice constant, and by the limitations
imposed by the momentum truncation for the correlator matrices.
This situation makes it difficult to reliably extract details of $V(r)$.
(The current lattice parameters should be better suited for
studying interactions involving, larger sized, baryons.)
In addition to $V(r)$ there is also present a
nonlocal potential \cite{Fie98b} which has not yet been computed.
 
The scattering phase shifts obtained from the underlying
lattice study show repulsive behavior in the low-momentum region and,
in this respect, compare favorably to experimental findings.
Quantitatively, the computed phase shifts are too small
by a sizeable factor. Relativistic corrections are at the 40\% level at
$p\approx 0.6a^{-1}$.
\vspace{1ex}

Acknowledgement: We would like to thank R.M. Woloshyn for a multiple-mass
solver.

\end{document}